\documentclass[11pt]{article}

\usepackage{amsmath}
\usepackage{graphicx}
\usepackage{amsfonts}
\usepackage{amssymb}
\usepackage{epsfig}
\usepackage{color}
\usepackage{psfrag}
\usepackage{epstopdf}

\newcommand{\EQ}{\begin{equation}}
\newcommand{\EN}{\end{equation}}
\newcommand{\be}{\begin{equation}}
\newcommand{\ee}{\end{equation}}
\newcommand{\bea}{\begin{eqnarray}}
\newcommand{\eea}{\end{eqnarray}}

\begin{document} \setcounter{page}{0}
\topmargin 0pt
\oddsidemargin 5mm
\renewcommand{\thefootnote}{\arabic{footnote}}
\newpage
\setcounter{page}{0}
\topmargin 0pt
\oddsidemargin 5mm
\renewcommand{\thefootnote}{\arabic{footnote}}
\newpage
\begin{titlepage}
\begin{flushright}
\end{flushright}
\vspace{0.5cm}
\begin{center}
{\large {\bf Classifying Potts critical lines}}\\
\vspace{1.8cm}
{\large Gesualdo Delfino$^{1,2}$ and Elena Tartaglia$^{1,2}$}\\
\vspace{0.5cm}
{\em $^1$SISSA -- Via Bonomea 265, 34136 Trieste, Italy}\\
{\em $^2$INFN sezione di Trieste}\\
\end{center}
\vspace{1.2cm}

\renewcommand{\thefootnote}{\arabic{footnote}}
\setcounter{footnote}{0}

\begin{abstract}
\noindent
We use scale invariant scattering theory to exactly determine the lines of renormalization group fixed points invariant under the permutational symmetry $S_q$ in two dimensions, and show how one of these scattering solutions describes the ferromagnetic and square lattice antiferromagnetic critical lines of the $q$-state Potts model. Other solutions we determine should correspond to new critical lines. In particular, we obtain that a $S_q$-invariant fixed point can be found up to the maximal value $q=(7+\sqrt{17})/2$. This is larger than the usually assumed maximal value 4 and leaves room for a second order antiferromagnetic transition at $q=5$. 
\end{abstract}
\end{titlepage}

\newpage

\section{Introduction}
Symmetry plays a prominent role within the theory of critical phenomena. The circumstance is usually illustrated referring to ferromagnetism, for which systems with different microscopic realizations but sharing invariance under transformations of the same group $G$ of internal symmetry fall within the same universality class of critical behavior. In the language of the renormalization group (see e.g. \cite{Cardy_book}) this amounts to say that the critical behavior of these ferromagnets is ruled by the same $G$-invariant fixed point. In general, however, there are several $G$-invariant fixed points of the renormalization group in a given dimensionality. Even staying within ferromagnetism, a system with several tunable parameters may exhibit multicriticality corresponding to fixed points with the same symmetry but different field content. 

The nontrivial relation between symmetry and criticality is further appreciated when extending the discussion to antiferromagnets. In this case, given the symmetry group $G$ under which the Hamiltonian is invariant, the critical exponents that the system exhibits {\it do} depend on microscopic details such as lattice structure and inclusion of next to nearest neighbor interactions. This forces a case by case analysis and, in principle, leaves room for a variety of $G$-invariant renormalization group fixed points. 

Remarkably, symmetry proves essential also in the theoretical study of ``geometrical'' critical phenomena which display no symmetry at all. A well known example is that of the percolation transition \cite{SA}, that is studied analytically as the limit $q\to 1$ of the ferromagnetic transition of the $q$-state Potts model \cite{Wu}, characterized by permutational symmetry $S_q$. This example is sufficient to understand the interest of enlarging the perspective to symmetries depending on a parameter that can be made continuous. In principle this allows to follow the evolution of critical behavior along lines of fixed points obtained varying the parameter, and to see for which parameter intervals criticality can be achieved. 

When aiming at the investigation of the relation between symmetry and criticality, the two-dimensional case is singled out for two main reasons. In the first place, since the room for nontrivial fixed points increases as the dimensionality decreases, the two-dimensional case offers the largest spectrum. The second reason is that in two dimensions the infinite-dimensional nature of the conformal group entitles to look for an exact description of fixed points, and conformal field theory has indeed provided a large amount of results extending to multipoint correlation functions \cite{DfMS}. However, some desirable pieces of the picture are still missing, a more global insight into the problem of antiferromagnets being one of them. Also, conformal field theory of geometrical criticality continues to be an open problem which is again well illustrated by percolation: it proved so far quite difficult to extend to real values of $q$ the $S_q$-invariant conformal theories known for $q=2,3,4$. As a consequence, for percolation on the whole plane, exact universal results have been longtime limited to critical exponents \cite{Nienhuis,DF}. Only more recently a basic quantity such as three-point cluster connectivity has been exactly predicted \cite{DV_3point} and numerically confirmed \cite{ZSK,PSVD}. 

In this paper we obtain global properties of two-dimensional renormalization group fixed points characterized by $S_q$ symmetry, for $q$ real. We do this within the framework of scale invariant scattering theory \cite{paraf} in which one studies the interaction among the particles of the underlying (1+1)-dimensional relativistic quantum theory\footnote{See \cite{fpu} for a perspective on particles, fields and critical phenomena in two dimensions.}. This approach uses infinite-dimensional conformal invariance in a quite indirect way, namely as the condition ensuring complete elasticity, and then exact solvability, of the scattering problem. Scale invariance is then sufficient to select scattering solutions corresponding to fixed points. An additional feature of the approach is that symmetry is manifest from the very beginning in the particle basis, and that analytic continuation in the symmetry parameter is naturally implemented. This allows us to obtain the $S_q$-invariant lines of fixed points parametrized by $q$, as well as the range of $q$ in which each of them is defined. The number of these fixed lines (which are listed in Table~\ref{solutions} below) is relatively small and may appear insufficient to account for the potential variety of $S_q$-invariant critical behavior. However, a different picture emerges from the theory and from comparison with known results. The critical properties of $S_q$-invariant systems are studied on the lattice through the $q$-state Potts model. For the latter, early exact calculations \cite{Baxter,Baxter_square_AF,BTA,Baxter_triangular} allowed to identify second order transition lines spanning the range $q\in[0,4]$, both in the ferromagnetic and antiferromagnetic case. The ferromagnetic transition is known to become first order\footnote{First order transitions are characterized by a finite correlation length and do not correspond to renormalization group fixed points. Throughout the paper, unless otherwise stated, our use of the term {\it critical} refers to second order criticality.} for $q>4$, and second order criticality is normally considered to be confined to $q\leq 4$ also for antiferromagnets. One of our scattering solutions spans the interval $q\in[0,4]$ and we will show how it is able to account for the known critical lines (critical and tricritical ferromagnetic lines, square lattice antiferromagnetic line) and possibly others in the same range. On the other hand, we find lines of fixed points which are defined in other ranges and should correspond to new second order transition lines. In particular, we find that a $S_q$-invariant fixed point can be found up to $q_{max}=(7+\sqrt{17})/2=5.5615..$. We will argue that available numerical results for Potts antiferromagnets leave room for new fixed points in the range $4<q\leq q_{max}$.

The mechanism through which the same scattering solution accounts for critical lines with different exponents is interesting and generalizes beyond the case of $S_q$ symmetry. The canonical critical exponents are determined by the conformal dimensions $\Delta_\sigma$ and $\Delta_\varepsilon$ of the order parameter field $\sigma$ and of the energy density field $\varepsilon$. Scale invariant scattering theory, however, brings forward the role of a third field, namely the chiral field $\eta$ which creates the massless particles. A scattering solution determines the dimension of $\eta$, in a way which is not unique. Different choices of $\Delta_\eta$ will correspond to different dimensions for $\sigma$ and $\varepsilon$, and then to different critical exponents. We also observe how the triad $\varepsilon$, $\sigma$, $\eta$ supports a form of duality that we call the $\varepsilon$-$\eta$ duality: given a critical line characterized by $\Delta_\varepsilon$, $\Delta_\sigma$ and $\Delta_\eta$, a new critical line with the same $\Delta_\sigma$ and different symmetry corresponds to the interchange of $\Delta_\varepsilon$ and $\Delta_\eta$. 

It is worth stating explicitly that the study of fixed points with permutational symmetry that we perform in this paper does not include random fixed points, that is fixed points of systems with quenched disorder. It has recently been shown \cite{random} how scale invariant scattering theory for a group $G\times S_n$ gives exact access (for $n\to 0$) to $G$-invariant random fixed points.  The corresponding space of solutions for the case $G=S_q$ will be given in \cite{DT}.

The paper is organized as follows. In the next section we recall the scale invariant scattering formalism and explain the origin of $\varepsilon$-$\eta$ duality, before deriving the $S_q$-invariant scattering solutions in section~3. The relation with Potts critical lines is obtained in section~4 for the critical and tricritical ferromagnet and the square lattice antiferromagnet. Section~5 contains some final remarks, in particular about existing numerical results for the antiferromagnetic case.

\section{Particles and fields at criticality}
A Euclidean field theory in two space dimensions corresponds to the continuation to imaginary time of a relativistic quantum field theory with one space and one time dimension. With this in mind, we switch from one case to the other according to convenience. Renormalization group fixed points of ($1+1$)-dimensional quantum field theories possess specific features at the level of the particle description. The first is that the particles are right or left movers with energy and momentum related as $p=e>0$ and $p=-e<0$, respectively. The second is that infinite-dimensional conformal symmetry implies that the scattering has to preserve infinitely many conserved quantities and is then completely elastic (the final state is kinematically identical to the initial one). As a consequence, the scattering amplitude of a right mover with a left mover can only depend on the center of mass energy, which is the only relativistic invariant in the process. This invariant, however, is dimensionful, so that scale invariance and unitarity imply that the amplitude is a constant. It follows that, denoting by $A_a(p)$ a particle of species $a$ with momentum $p$, and by $S_{ab}^{cd}$ the amplitude for the process with inital state $A_a(p_1)A_b(p_2)$ and final state $A_c(p_1)A_d(p_2)$, the unitarity and crossing equations \cite{ELOP} for right-left scattering take the particularly simple form \cite{paraf,fpu}
\EQ
\sum_{e,f}S_{ab}^{ef}\left[S_{ef}^{cd}\right]^*=\delta_a^c\delta_b^d\,,
\label{unitarity}
\EN
\EQ
S_{ab}^{cd}=\left[S_{a\bar{d}}^{c\bar{b}}\right]^*,
\label{crossing}
\EN
where the asterisk denotes complex conjugation and the bar over indices denotes antiparticles.

The energy independence of the amplitudes expresses the fact that they actually encode the statistical properties associated to position exchange of the particles on the line; a relation with the spin of the fields which create the particles is then expected. We recall that a field $\Phi(x)$ with conformal dimensions $(\Delta_\Phi,\bar{\Delta}_\Phi)$ has scaling dimension 
\EQ
X_\Phi=\Delta_\Phi+\bar{\Delta}_\Phi
\EN
and spin 
\EQ
s_\Phi=\Delta_\Phi-\bar{\Delta}_\Phi\,.
\EN
Right and left movers are created by chiral fields $\eta$ and $\bar{\eta}$ with conformal dimensions $(\Delta_\eta,0)$ and $(0,{\Delta}_\eta)$, respectively. Upon diagonalization of the scattering in the neutral channel, the scattering eigenvalue $S$, which by unitarity is a phase, is related to $s_\eta=\Delta_\eta$ as \cite{paraf,fpu}
\EQ
S=e^{-2i\pi\Delta_\eta}\,.
\label{phase}
\EN
Bosons and fermions are obtained for $\Delta_\eta$ integer and half-integer, respectively, and particles with generalized statistics otherwise\footnote{Away from criticality, generalized statistics does not enter low energy applications (see e.g. \cite{wedge}), but appears in high energy asymptotics \cite{Smirnov,sqf,DN_ttbar}.}.

As the other physically interesting fields of the theory, $\eta$ is local with respect to the energy density field $\varepsilon$, which is the most relevant (i.e. with smallest scaling dimension) spinless field left invariant by the internal symmetry (excluding the identity). In general, two fields $\Phi_i$ and $\Phi_j$ are said to be mutually local if a correlator $\langle\cdots\Phi_i(x)\Phi_j(0)\cdots\rangle$ is single valued when $x$ is taken around $0$. For statistical mechanical applications we think of $x=(x_1,x_2)$ as the coordinate on the Euclidean plane. The operator product expansion (OPE) 
\EQ
\Phi_i(x)\Phi_j(0)=\sum_k C_{ij}^{k}\,z^{\Delta_{k}-\Delta_{i}-\Delta_{j}}\bar{z}^{\bar{\Delta}_{k}-\bar{\Delta}_{i}-\bar{\Delta}_{j}}\,\Phi_k(0)\,,
\label{ope}
\EN
where $\Delta_i\equiv\Delta_{\Phi_i}$, $z=x_1+ix_2$ and $\bar{z}=x_1-ix_2$, allows to write the mutual locality condition (invariance under $z\to e^{2i\pi}z$, $\bar{z}\to e^{-2i\pi}\bar{z}$) as 
\begin{equation}
s_{\Phi_i} + s_{\Phi_j} - s_{\Phi_k} \in \mathbb{Z}\,;
\label{gamma}
\end{equation}
normally the fields $\Phi_k$ contributing to the r.h.s. of (\ref{ope}) have spins which differ by integers, so that it is sufficient to check (\ref{gamma}) for one of them. Since $s_\varepsilon=0$ and the OPE $\varepsilon\cdot\eta$ produces fields $\Phi_k$ with $\bar{\Delta}_k=\Delta_\varepsilon$, the condition of mutual locality between $\varepsilon$ and $\eta$ reads
\EQ
\Delta_\eta-\Delta_{k}+\Delta_\varepsilon\in\mathbb{Z}\,.
\label{duality}
\EN

Concerning the order parameter field $\sigma(x)$ ($s_\sigma=0$), it carries a representation of the internal symmetry and its OPE with the energy density is of the form
\EQ
\sigma\cdot\varepsilon\sim\sigma+\ldots\,.
\EN
Here and in the following we write OPEs omitting coefficients and coordinate dependence, giving for granted that the complete form follows from (\ref{ope}). Generically, $\sigma$ is nonlocal with respect to $\eta$, a property encoded by an OPE of the form
\EQ
\sigma\cdot\eta\sim\tilde{\sigma}+\ldots\,,
\label{sigma_eta}
\EN
with $s_{\tilde{\sigma}}=0$ (no mutual locality for $\Delta_\eta$ noninteger). On the other hand, $\bar{\Delta}_\eta=0$ implies $\bar{\Delta}_{\tilde{\sigma}}=\bar{\Delta}_\sigma$. Hence, $\sigma$ and ${\tilde{\sigma}}$ have identical conformal dimensions, but are mutually nonlocal and cannot coincide. This structure is well known for the Ising model, where $\eta$ is a free fermion and ${\tilde{\sigma}}$ the disorder field associated to high-low temperature duality. The structure, however, is more general and extends also to the case of continuous symmetries, which in two dimensions do not break spontaneously and do not give rise to the canonical order-disorder duality. 

A remarkable property of the relations (\ref{duality})-(\ref{sigma_eta}) is their symmetry under the exchange of $\Delta_\varepsilon$ with $\Delta_\eta$. Given a critical point with $\Delta_\varepsilon=\Delta_1$, $\Delta_\eta=\Delta_2$ and a $\Delta_\sigma$, also a critical point with $\Delta_\varepsilon=\Delta_2$, $\Delta_\eta=\Delta_1$ and the same $\Delta_\sigma$ will satisfy the relations. We will see explicit examples of this ``$\varepsilon$-$\eta$ duality'' in section 4. 

\begin{figure}
\begin{center}
\includegraphics[width=10cm]{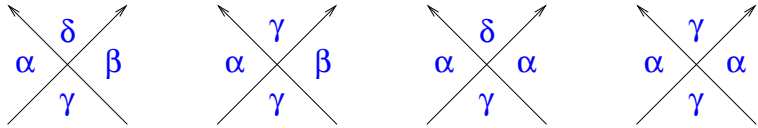}
\caption{Amplitudes $S_0$, $S_1$, $S_2$ and $S_3$ of the $S_q$-invariant theory. Different letters correspond to different colors.}
\label{potts_ampl}
\end{center} 
\end{figure}

\section{$S_q$-invariant scattering solutions}
We now implement the scattering approach of the previous section for the case of scale invariant theories possessing permutational symmetry $S_q$. The first step is to introduce a particle basis carrying a representation of the symmetry. For $S_q$ this is achieved considering particles $A_{\alpha\beta}$ with $\alpha,\beta=1,2,\dots,q$, and $\alpha\neq\beta$. For the Potts ferromagnet below critical temperature these particles correspond to the kinks that interpolate between pairs of the $q$ degenerate ground states \cite{CZ}. It was argued in \cite{paraf,random} that this particle basis has to be identified as the fundamental way of representing $S_q$ symmetry also at criticality (where the ground states coalesce and there are no kinks\footnote{See however \cite{KA}, where it was shown that domain wall configurations play an important role for finite size corrections in the critical Ising ferromagnet. We thank an anonymous referee for bringing these papers to our attention.}) and beyond the ferromagnetic case. The results of the present paper will provide an illustration of this point. 

\begin{figure}
\begin{center}
\includegraphics[width=10cm]{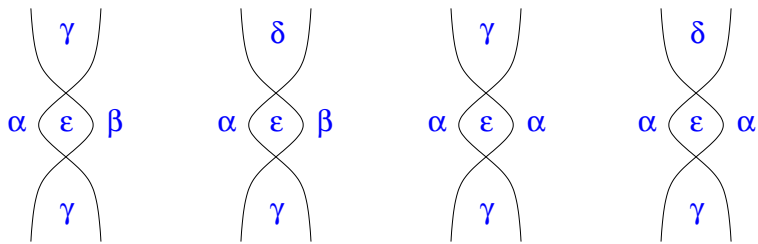}
\caption{Pictorial representations associated to the unitarity equations (\ref{unit1}), (\ref{unit2}), (\ref{unit3}), (\ref{unit4}), in that order. The amplitude for the lower crossing multiplies the complex conjugate of the amplitude for the upper crossing, and sum over $\varepsilon$ is implied.}
\label{uni_diagrams}
\end{center} 
\end{figure}

In general, we think of the trajectory of the particle $A_{\alpha\beta}$ as a line separating a region of the plane characterized by the value (``color") $\alpha$ from a region characterized by the color $\beta$. Permutational invariance then yields the four inequivalent amplitudes $S_0$, $S_1$, $S_2$ and $S_3$ depicted in Fig.~1. For these the crossing relations (\ref{crossing}) yield
\EQ
S_0=S_0^*\equiv\rho_0\,,\hspace{1cm}S_1=S_2^*\equiv\rho e^{i\varphi}\,,\hspace{1cm}S_3=S_3^*\equiv\rho_3\,,
\EN
where we introduced 
\EQ
\rho\geq 0\,,\hspace{1cm}\rho_0,\,\rho_3,\,\varphi\in\mathbb{R}\,.
\label{reality}
\EN
With this parametrization the unitarity equations (\ref{unitarity}) translate into (see also Fig.~\ref{uni_diagrams})
\bea
&&(q-3)\rho_0^2+\rho^2=1\,,\label{unit1}\\
&&(q-4)\rho_0^2+2\rho_0\rho\cos\varphi=0\,,\label{unit2}\\
&&(q-2)\rho^2+\rho_3^2=1\,,\label{unit3}\\
&&(q-3)\rho^2+2\rho\rho_3\cos\varphi=0\,.\label{unit4}
\eea
Note that the equations contain $q$ as a parameter which does not need to be integer, so that the scattering formalism realizes in the continuum the analytic continuation in $q$ which is known from the lattice representation of the symmetry (see next section). For $q$ integer, the amplitudes involving a number of colors larger than that integer ($S_0$ for $q=3$, also $S_1$ and $S_2$ for $q=2$) are unphysical. All amplitudes, however, enter the continuation to non-integer values of $q$. Various mechanisms of this continuation, in particular for the case $q\to 1$ relevant for percolation, are illustrated in \cite{DV_3point,DVC,DV_4point,DV_crossing,DG_confinement}. 

The solutions of the unitarity equations are listed in Table~\ref{solutions} together with the range of $q$ in which they satisfy the conditions (\ref{reality}). The sign doublings follow from the general fact that, given a solution of the unitarity and crossing equations (\ref{unitarity}) and (\ref{crossing}), another solution is obtained reversing the sign of all amplitudes. In our notation, II$_+$ (II$_-$) corresponds to the solution with upper (lower) signs, and similarly for III, IV and V. 

Solution I, which has $\varphi$ as a free parameter, is defined for $q=3$ only. Although $S_0$ is unphysical at $q=3$, we quote the values of $\rho_0$ allowed by the equations for the purpose of comparison with solutions II, III and IV, which allow continuation away from $q=3$. 

We conclude this section observing that it follows in general from the amplitudes of Fig.~\ref{potts_ampl} that the state $\sum_{\gamma\neq\alpha}A_{\alpha\gamma}A_{\gamma\alpha}$ scatters into itself with the amplitude
\EQ
S=S_3 + (q-2)S_2\,,
\label{eigenvalue}
\EN
which is the phase entering (\ref{phase}).

\begin{table}
\begin{center}
\begin{tabular}{|c|c||c|c|c|c|}
\hline
Solution & Range & $\rho_0$ & $\rho$ & $2\cos\varphi$ & $\rho_3$ \\
\hline
I & $q=3$ &$0$, $2\cos\varphi$ & $1$ & $\in[-2,2]$ & $0$ \\ 
& & & & & \\
II$_\pm$ & $q\in[-1,3]$ & $0$ & $1$ & $\pm\sqrt{3-q}$ & $\pm \sqrt{3-q}$\\
& & & & & \\
III$_\pm$ & $q\in[0,4]$ & $\pm 1$ & $\sqrt{4-q}$ & $\pm\sqrt{4-q}$ & $\pm (3-q)$\\
& & & & & \\
IV$_\pm$ & $q\in[\frac{1}{2}(7-\sqrt{17}),3]$ & $\pm \sqrt{\frac{q-3}{q^2-5q+5}}$ & $\sqrt{\frac{q-4}{q^2-5q+5}}$ & $\pm\sqrt{(3-q)(4-q)}$ & $\pm \sqrt{\frac{q-3}{q^2-5q+5}}$\\
& & & & & \\
V$_\pm$ & $q\in[4,\frac{1}{2}(7+\sqrt{17})]$ & $\pm \sqrt{\frac{q-3}{q^2-5q+5}}$ & $\sqrt{\frac{q-4}{q^2-5q+5}}$ & $\mp\sqrt{(3-q)(4-q)}$ & $\pm \sqrt{\frac{q-3}{q^2-5q+5}}$\\
\hline
\end{tabular}
\caption{Solutions of Eqs.~(\ref{unit1})-(\ref{unit4}) with the conditions (\ref{reality}). They correspond to renormalization group fixed points of $S_q$-invariant theories.} 
\label{solutions}
\end{center}
\end{table}

\section{Potts critical lines}
\subsection{Potts model and Fortuin-Kasteleyn representation}
The $q$-state Potts model \cite{Wu} is defined on the lattice by the Hamiltonian 
\EQ
{\cal H}=-J\sum_{\langle i,j\rangle}\delta_{s_i,s_j}\,,\hspace{1cm}s_i=1,2,\ldots,q\,,
\label{potts}
\EN
where the sum is taken over nearest neighboring sites. The model generalizes the Ising model ($q=2$) to the case in which the site variable takes $q$ colors, and is clearly invariant under permutations of the colors. 

An important feature of the $q$-state Potts model is that the partition function admits the graph (or Fortuin-Kasteleyn) expansion \cite{FK}
\EQ
Z\equiv\sum_{\{s_i\}}e^{-{\cal H}/T}\propto\sum_G p^{N_b}(1-p)^{\bar{N}_b}q^{N_c}\,,
\label{FK}
\EN
where $G$ is a graph obtained placing bonds on the edges of the lattice, $N_b$ is the number of bonds in $G$, $\bar{N}_b$ the number of edges without a bond, and 
\EQ
p=1-e^{-J/T}\,,
\label{p}
\EN
$T$ being the temperature; $N_c$ is the number of clusters in $G$, with a cluster corresponding to a set of connected bonds, but also to a site not touched by any bond. This representation of the model is also known as random cluster model, since it corresponds to configurations which are obtained placing bonds with probability $p$; each of the resulting clusters can take $q$ different colors. Within this representation $q$ appears as a parameter which is no longer restricted to take integer values. In particular, for $q\to 1$ the weight $p^{N_b}(1-p)^{\bar{N}_b}$ of a bond configuration coincides with that of the percolation problem, in which edges are randomly occupied with probability $p$ and color plays no role. Percolation is characterized by the existence of a critical value $p_c$ above which an infinite cluster is found with non-zero probability. Equation (\ref{FK}) relates such a percolation transition to the spontaneous symmetry breaking of $S_q$ symmetry in the $q\to 1$ Potts ferromagnet. The Fortuin-Kasteleyn random cluster representation makes sense of the Potts model with non-integer $q$ also in the antiferromagnetic case ($J<0$), in spite of the absence of a probabilistic interpretation\footnote{Notice, on the other hand, that for $T\to 0$ the partition function of the Potts antiferromagnet counts the number of ways the sites of a lattice can be colored with $q$ colors in such a way that nearest neighbors always have different colors ($q$-coloring problem, see \cite{Wu}).} ($p<0$). In the following, when talking of the Potts model, we will generally understand its continuation to real values of $q$. 

In the two-dimensional case, a renormalization group fixed point of the Potts model should correspond to one of the $S_q$-invariant fixed points identified in the previous section within the scale invariant scattering formalism. The following subsections explore this correspondence.

\subsection{Ferromagnetic critical line}
Since the ferromagnetic phase transition in the two-dimensional $q$-state Potts model is known to be of the second order up to $q=4$ \cite{Baxter}, the critical ferromagnetic line must correspond to one of the solutions in Table~\ref{solutions} having $q=4$ as upper endpoint. The fact that in two dimensions the Ising model is a theory of free fermions implies $\rho_3=-1$, and uniquely selects the solution III$_-$. This identification was already obtained in \cite{paraf} and we now recall how it is further characterized in the language of conformal field theory. 

Two-dimensional conformal field theories \cite{DfMS} are first of all characterized by the central charge $c$, which grows with the number of degrees of freedom of the system. The four-state Potts model is a particular case of the Ashkin-Teller model (see \cite{DG} for the scattering description), which corresponds to two Ising models coupled by a four spin interaction and has $c=1$. The critical line III$_-$ must then be able to account for a conformal field theory with central charge $c(q)\leq 1$. In this range of central charge a main physical role is played by the so-called ``degenerate'' primary fields $\Phi_{m,n}(z)$  \cite{BPZ} with conformal dimension
\EQ
\Delta_{m,n}=\frac{[(p+1)m-pn]^2-1}{4p(p+1)}\,,
\label{deltamunu}
\EN
where $m,n=1,2,\ldots$, and $p$ parametrizes the central charge as
\EQ
c=1-\frac{6}{p(p+1)}\,;
\label{cc}
\EN
one similarly has $\bar{\Phi}_{m,n}(\bar{z})$ with dimension $\bar{\Delta}_{m,n}$, and a complete degenerate field is the product of a $z$ and a $\bar{z}$ part, with different indices if the spin is non-zero. The OPE is specified by
\EQ
\Phi_{m_1,n_1}\cdot\Phi_{m_2,n_2}\sim\sum_{k=0}^{min(m_1,m_2)-1}\,\sum_{l=0}^{min(n_1,n_2)-1}\left[\Phi_{|m_1-m_2|+1+2k,|n_1-n_2|+1+2l}\right]\,,
\label{dd}
\EN
together with a similar relation for the fields $\bar{\Phi}_{m,n}$.

The strategy for relating this ``colorless'' conformal field theory to the $S_q$-invariant scattering solution III$_-$ is the following (details are given in \cite{paraf}). It can be argued that the energy density field $\varepsilon(x)$ for the ferromagnetic model is a degenerate field. Then, knowing that $\Delta_\varepsilon=1/2$ at the Ising point ($c=1/2$, $p=3$), that $\varepsilon$ cannot produce more relevant fields in the OPE with itself, and that it is odd under the high-low temperature duality characteristic of the model, one arrives at the identification $\Delta_\varepsilon=\Delta_{2,1}$. One then looks for the particle-creating field $\eta$ as the most relevant chiral field local with respect to $\varepsilon$. It can be shown that also $\eta$ needs to be degenerate, so that (\ref{dd}) can be used to obtain $\Delta_\eta=\Delta_{1,3}$. This result for $\Delta_\eta$ as a function of $p$ can then be compared with that provided by (\ref{phase}) as a function of $q$ (Eq.~(\ref{eigenvalue}) gives $S=\mp e^{-4i\varphi}$ for the solutions III$_\pm$). This provides the relation
\EQ
\sqrt{q}=2\sin\frac{\pi(p-1)}{2(p+1)}=2\cos\frac{\pi}{p+1}\,,
\label{q_cr}
\EN
which determines the central charge as a function of $q$. This result coincides with that obtained in \cite{DF} from the exact lattice determination of scaling dimensions \cite{Nienhuis}; here it is derived in a self-contained way in the continuum limit. A slightly more general analysis involving nondegenerate fields \cite{paraf} allows to find $\Delta_\sigma=\Delta_{1/2,0}$. 

The $\varepsilon$-$\eta$ duality associated to (\ref{duality})-(\ref{sigma_eta}) relates the Potts ferromagnetic critical line to a critical line with $\Delta_\varepsilon=\Delta_{1,3}$ and $\Delta_\eta=\Delta_{2,1}$. The latter corresponds to the scale invariant scattering solution with $O(n)$ symmetry, $n=2\cos\frac{\pi}{p}$ \cite{paraf}. 

\subsection{Ferromagnetic tricritical line}
When the Potts Hamiltonian (\ref{potts}) is generalized allowing for the presence of vacant sites, tricriticality can be realized. The tricritical line exists as long as the critical one exists, and the two lines meet at the common endpoint $q=4$ \cite{NBRS}. The presence of vacancies preserves color permutational symmetry, so that the tricritical line must also correspond to one of the scattering solutions of section~3. Since the solutions III$_\pm$ are the only ones terminating at $q=4$, and since they do not coincide at $q=4$, we are again left with III$_-$ as the only possibility. 

Hence, besides that of the previous subsection, there should be another relation between solution III$_-$ and conformal field theory with $c\leq 1$, a relation  corresponding to the tricritical line. This is indeed found as follows. The energy density field on the tricritical line must have the same OPE and duality properties as on the critical line. We also know the value of $\Delta_\varepsilon$ at $q=4$ where the two lines meet. This information then selects $\Delta_\varepsilon=\Delta_{1,2}$. Since this differs from the result on the critical line by exchange of the two indices, the form of the OPE (\ref{dd}) ensures that the search for $\eta$ as a chiral field local with respect to $\varepsilon$ has a solution with the same exchange, i.e. $\Delta_\eta=\Delta_{3,1}$; one can check that this is indeed the most relevant solution. We can now use this result in (\ref{phase}), with $S=e^{-4i\varphi}$ for solution III$_-$, to obtain
\begin{equation}
\sqrt{q} = 2\sin\frac{\pi(p-2)}{2p}=2\cos\frac{\pi}{p}\,;
\end{equation}
comparison with (\ref{q_cr}) shows that the same $q$ corresponds to $p$ on the critical line and to $p+1$ on the tricritical one. For the order parameter one obtains $\Delta_\sigma=\Delta_{0,1/2}$. Again, these findings coincide with those of \cite{DF,Nienhuis}. $\varepsilon$-$\eta$ duality relates this tricritical line to the scattering solution with $O(n)$ symmetry, $n=2\cos\frac{\pi}{p+1}$ \cite{paraf}.

\subsection{Critical line at $q=3$}
Solution I contains $\varphi$ as a free parameter and then corresponds to a line of fixed points with $q=3$. Such a line is generated by a truly marginal field and does not fit within the basic lattice realization (\ref{potts}) of $S_3$ symmetry. Indeed, in two-dimensional spin models with discrete internal symmetry the presence of a line of fixed points requires an additional interaction parameter besides the exchange coupling $J$ (see e.g. the Ashkin-Teller model \cite{DG}). The Hamiltonian (\ref{potts}), however, can describe a point on this line, and an explicit example of this situation will be given in the next subsection. For the time being we explain why such a line arises at $q=3$. The point is that $S_3$ can be realized by cyclic $\mathbb{Z}_3$ permutations together with a $\mathbb{Z}_2$ reflection. The particles admit the identifications $A_{\alpha,\alpha+1(\textrm{mod}\,3)}\equiv A$, $A_{\alpha,\alpha-1(\textrm{mod}\,3)}\equiv\bar{A}$, where $A$ and $\bar{A}$ carry $\mathbb{Z}_3$ charges $1(\textrm{mod}\,3)$ and $-1(\textrm{mod}\,3)$, respectively, and are exchanged by the $\mathbb{Z}_2$ reflection (charge conjugation). In general, however, a doublet of particles $A$ and $\bar{A}$ with opposite charges can represent a symmetry $U(1)$. It is well known that the minimal realization of this symmetry in two-dimensional conformal field theory is provided by the free bosonic action
\EQ
{\cal A}=\frac{1}{4\pi}\int d^2x\,(\partial_a\phi)^2\,,
\label{action}
\EN
which indeed describes a line of fixed points with central charge $c=1$. The energy density field $\varepsilon(x)=\cos 2b\phi(x)$, with scaling dimension $X_\varepsilon=2b^2$, contains the parameter $b$ which provides the coordinate along the fixed line. It was shown in \cite{paraf} that $\Delta_\eta=1/4b^2$ on this line. Since (\ref{eigenvalue}) gives $S=S_2=e^{-i\varphi}$ for solution I, (\ref{phase}) yields
\EQ
\varphi=\frac{\pi}{2b^2}
\label{phi_b}
\EN
for the scattering along the line. As required, $S=-1$ at the point $b^2=1/2$ where (\ref{action}) also admits a free fermion representation (see \cite{fpu} and references therein). $U(1)$ symmetry also yields $\Delta_\sigma=1/16b^2$ (see \cite{paraf}).

\subsection{Critical line of the square lattice antiferromagnet} 
The Potts model (\ref{potts}) can become critical also in the antiferromagnetic case ($J<0$). An antiferromagnet tries to find at low temperature a ground state in which nearest neighboring spins take different values. The number of such configurations can be zero, finite or infinite depending on the lattice. It follows that, in contrast with the universality exhibited by ferromagnets, antiferromagnetic critical behavior essentially depends on the lattice structure and needs to be investigated case by case. In this subsection we discuss the implications of our continuum approach for the case which is best understood in the discrete formulation \cite{Baxter,Baxter_square_AF}, that of the square lattice.

The $q=2$ (Ising) square lattice antiferromagnet has two ground states: one in which all spins on the even sublattice have color 1 and the spins on the odd sublattice have color 2, and one in which the even sublattice has color 2 and the odd sublattice has color 1. The phase transition is essentially the same as in the ferromagnet, with the order parameter now corresponding to the staggered magnetization, which differs from the usual one by a multiplicative factor $(-1)^P$, $P$ being the sublattice parity. In particular, for the central charge and the scaling dimensions at the critical point, one still has $c=\Delta_\varepsilon=\Delta_\eta=1/2$, and $\Delta_{\sigma}=1/8$, with $\Delta_\sigma$ referring to the staggered magnetization in the antiferromagnetic case. 

For $q=3$ the number of antiferromagnetic ground states on the square lattice is infinite, and the $T=0$ case can be mapped onto a specific case of the six-vertex model \cite{Baxter,LW,BH}, which is known to be critical and to renormalize on the $c=1$ conformal field theory with action (\ref{action}). This means that, within our scattering classification, this $T=0$ critical point corresponds to a point of solution I identified by a specific value of $b^2$ in (\ref{phi_b}). This value can be determined recalling that Baxter showed that the square lattice Potts antiferromagnet has a second order transition for $q\in[0,4]$, although for $q>3$ this no longer corresponds to physical values of the temperature \cite{Baxter_square_AF}. The range of existence of this transition selects the scattering solutions III$_\pm$, and the fact that $\Delta_\eta=1/2$ at $q=2$ finally identifies solution III$_-$. The latter has $2\cos\varphi=-1$ at $q=3$, and this yields\footnote{Note that $\varphi=4\pi/3$ would give $b^2=3/8$. However, it was shown in \cite{AF1,AF2} that the interesting range in this case is $b^2>1/2$ (repulsive regime for $T\to 0$).} $b^2=3/4$ through (\ref{phi_b}). This in turn gives $\Delta_\varepsilon=b^2=3/4$, in agreement with the lattice determination of \cite{CJS}. 

It was argued in \cite{AF2} that in presence of temperature and staggered polarization the $q=3$ model exhibits a phase transition associated to a renormalization group flow from the $c=1$ antiferromagnetic fixed point to the $c=4/5$ ferromagnetic fixed point. It was also argued that this flow, later observed numerically in \cite{OO}, is the case $N=4$ of a family of integrable flows which was found in \cite{FZ_flows} to relate the $\mathbb{Z}_N$ parafermionic conformal field theories \cite{ZF_paraf} with central charge 
\EQ
c=\frac{2(N-1)}{N+2}
\label{c_N}
\EN
to the models with central charge (\ref{cc}) and $p=N+1$. For $N=2$ one consistently finds that the fixed points in the two series have the same central charge $1/2$, and it is natural to conjecture a general relation between the critical line of the square lattice $q$-state antiferromagnet and (\ref{c_N}). The relation between $q$ and $N$ was first obtained from the lattice by Saleur \cite{Saleur} (see also \cite{JS} for a very detailed study and \cite{Ikhlef} for a review of this approach). Within our framework this relation follows from the fact that, due to the $\varepsilon$-$\eta$ duality, the dimension $\Delta_\eta$ along the Potts antiferromagnetic line concides with that of the most relevant neutral field of the $\mathbb{Z}_N$ models (known from \cite{ZF_paraf}), and then reads $\Delta_\eta=2/(N+2)$. We can then use (\ref{phase}) with $S=e^{-4i\varphi}$ for solution III$_-$ to obtain
\EQ
q=4\cos^2\frac{\pi}{N+2}\,.
\label{q_N}
\EN
In turn, the dimension $\Delta_\varepsilon$ along the Potts antiferromagnetic line coincides with that of $\eta$ (fundamental parafermion) of the $\mathbb{Z}_N$ models (known from \cite{ZF_paraf}), and then reads $\Delta_\varepsilon=(N-1)/N$, in agreement with the lattice result of \cite{JS}. Concerning the order parameter of the antiferromagnet (staggered magnetization), its dimension should coincide with that of one of the order parameters of the $\mathbb{Z}_N$ models. These have dimensions $k(N-k)/2N(N+2)$, with $k=1,2.\ldots,N-1$ \cite{ZF_paraf}, and the desired values $1/16$ for $N=q=2$ and $1/12$ for $N=4$ ($q=3$) are obtained for $k=N/2$. This gives $\Delta_{\sigma}=N/8(N+2)$, in agreement with the lattice result of \cite{Saleur}. 

Some of the results of this section are summarized in Table~\ref{summary}.

\begin{table}
\begin{center}
\begin{tabular}{|c|c|c|c|c|c|c|}
\hline
$S_q$\,(III$_-$) & Potts & $c$ & $\Delta_{\varepsilon}$ & $\Delta_\eta$ & $\Delta_{\sigma}$ & $\varepsilon$-$\eta$ dual\\
\hline
$\sqrt{q}=2\cos\frac{\pi}{(p+1)}$ & F critical & $1-\frac{6}{p(p+1)}$ & $\Delta_{2,1}$ & $\Delta_{1,3}$ & $\Delta_{\frac{1}{2},0}$ & $O(n)$, $n=2\cos\frac{\pi}{p}$ \\
& & & & & & \\
$\sqrt{q}=2\cos\frac{\pi}{p}$ & F tricritical & $1-\frac{6}{p(p+1)}$ & $\Delta_{1,2}$ & $\Delta_{3,1}$ & $\Delta_{0,\frac{1}{2}}$ & $O(n)$, $n=2\cos\frac{\pi}{(p+1)}$ \\
& & & & & & \\
$\sqrt{q}=2\cos\frac{\pi}{(N+2)}$ & AF square & $\frac{2(N-1)}{N+2}$ & $\frac{N-1}{N}$ & $\frac{2}{N+2}$ & $\frac{N}{8(N+2)}$ & $\mathbb{Z}_N$ \\
\hline
\end{tabular}
\caption{Realizations of the $S_q$-invariant scattering solution III$_-$ as Potts ferromagnetic (F) and square lattice antiferromagnetic (AF square) critical lines. The central charge and the conformal dimensions of the energy density, chiral, and order parameter fields are given together with the critical lines obtained by $\varepsilon$-$\eta$ duality. The conformal dimensions $\Delta_{m,n}$ are specified through (\ref{deltamunu}).} 
\label{summary}
\end{center}
\end{table}

\section{Discussion}
It appears from the results collected in Table~\ref{summary} that the best known Potts critical lines all correspond to the scattering solution III$_-$. It is then natural to ask whether this solution possesses some special property. We can observe that it follows from (\ref{dd}) that the field $\eta$ of the critical and tricritical ferromagnetic lines ($\Phi_{1,3}$ and $\Phi_{3,1}$, respectively) satisfy the property\footnote{The OPE coefficient $C_{\eta\eta}^\eta$ vanishes at $q=2$.}
\EQ
 \eta\cdot\eta\sim I+\eta+\ldots\,,
\label{fusion}
\EN
which in terms of particles gives $A_{\alpha\gamma}A_{\gamma\beta}\sim A_{\alpha\beta}$. It was shown in \cite{paraf} that this relation yields a set of equations for the scattering amplitudes for which III$_-$ is the only solution. It does not seem true, however, that all critical lines described by III$_-$ satisfy (\ref{fusion}). Indeed, for the last case of Table~\ref{summary}, (\ref{fusion}) would amount by $\varepsilon$-$\eta$ duality to $\varepsilon\cdot\varepsilon\sim I+\varepsilon+\ldots$ in the $\mathbb{Z}_N$ models, a property which does not hold because $\varepsilon$ is odd under the high-low temperature duality of these models.  

We already recalled in the introduction that the exact lattice results for second order phase transitions in the Potts model refer to the interval $q\in[0,4]$. One can ask whether it is possible to find a second order phase transition in a Potts model with $q>4$. Our results indicate that this should be possible, as long as $q$ does not exceed $q_{max}=(7+\sqrt{17})/2=5.5615..$\,. Of course, since the ferromagnetic case is settled by universality, one should look to antiferromagnets. In this respect, it is interesting that infinite families of two-dimensional lattices allowing for an antiferromagnetic transition with $q$ arbitrarily large have been found in \cite{Huang}. The authors expect the transition to be first order whenever $q>4$, and verify numerically that this is the case for $q\gtrsim 8$. For $4<q\lesssim 8$ they state that the transition is presumably first order, with a large correlation length complicating numerical verification. These considerations seem to supersede the indication of \cite{DHJSS} that the transition observed numerically at $q=5$ on one of these lattices could be second order rather than weakly first order. It seems, however,  that the data do not yet exclude that this class of antiferromagnets can provide a realization of fixed points falling within our solution V.




\begin{thebibliography}{99}
\bibitem{Cardy_book} J. Cardy, Scaling and renormalization in statistical physics, Cambridge, 1996.
\bibitem{SA} D. Stauffer and A. Aharony, Introduction to Percolation Theory, 2nd edn, Taylor \& Francis, London, 1992.
\bibitem{Wu} F.Y. Wu, Rev. Mod. Phys. 54 (1982) 235.
\bibitem{DfMS} P. Di Francesco, P. Mathieu and D. Senechal, Conformal field theory, Springer-Verlag, New York, 1997.
\bibitem{Nienhuis} B. Nienhuis, Coulomb gas formulation of two-dimensional phase transitions, in C. Domb and J.L. Lebowitz (eds.), Phase transitions and critical phenomena, vol. 11, p. 1-53, Academic Press, London, 1987.
\bibitem{DF} Vl.S. Dotsenko and V.A. Fateev, Nucl. Phys. B 240 (1984) 312.
\bibitem{DV_3point} G. Delfino and J. Viti, J. Phys. A 44 (2011) 032001.
\bibitem{ZSK} R.M. Ziff, J.J.H. Simmons and Peter Kleban, J. Phys. A 44 (2011) 065002.
\bibitem{PSVD} M. Picco, R. Santachiara, J. Viti and G. Delfino, Nucl. Phys. B 875 (2013) 719.
\bibitem{paraf} G.~Delfino, Annals of Physics 333 (2013) 1.
\bibitem{fpu} G.~Delfino, Annals of Physics 360 (2015) 477.
\bibitem{Baxter} R.J. Baxter, Exactly Solved Models of Statistical Mechanics, Academic Press, London, 1982.
\bibitem{Baxter_square_AF} R.J. Baxter, Proc. Roy. Soc. London A 383 (1982) 43.
\bibitem{BTA} R.J. Baxter, H.N.V. Temperley and S.E. Ashley, Proc. R. Soc. Lond. A 358 (1978) 535.
\bibitem{Baxter_triangular} R.J. Baxter, J. Phys. A 19 (1986) 2821.
\bibitem{random} G.~Delfino, Phys. Rev. Lett. 118 (2017) 250601.
\bibitem{DT} G. Delfino and E. Tartaglia, On superuniversality in the q-state Potts model with quenched disorder, arXiv:1709.00364.
\bibitem{ELOP} R.J. Eden, P.V. Landshoff, D.I. Olive and J.C. Polkinghorne, The analytic S-matrix, Cambridge, 1966.
\bibitem{wedge} G. Delfino and A. Squarcini, Phys. Rev. Lett. 113 (2014) 
066101. 
\bibitem{Smirnov} F. Smirnov, Commun. Math. Phys. 132 (1990) 415.
\bibitem{sqf} G. Delfino, Nucl. Phys. B 807 (2009) 455.
\bibitem{DN_ttbar} G. Delfino and G. Niccoli, JHEP 05 (2006) 035. 
\bibitem{CZ} L. Chim and A.B. Zamolodchikov, Int. J. Mod. Phys. A 7 (1992) 5317.
\bibitem{KA} P. Kleban and G. Akinci, Phys. Rev. Lett. 51  (1983) 1058; Phys. Rev. B 28 (1983) 1466.
\bibitem{DVC} G. Delfino, J. Viti and J. Cardy, J. Phys. A 43 (2010) 152001.
\bibitem{DV_4point} G. Delfino and J. Viti, Nucl. Phys. B 852 (2011) 149.
\bibitem{DV_crossing} G. Delfino and J. Viti, J. Phys. A 45 (2012) 032005.
\bibitem{DG_confinement} G. Delfino and P. Grinza, Nucl. Phys. B 791 (2008) 265.
\bibitem{FK}  C.M. Fortuin and P.W. Kasteleyn, J. Phys. Soc. Jpn. Suppl. 26 (1969) 11; Physica 57 (1972) 536.
\bibitem{BPZ}  A.A. Belavin, A.M. Polyakov and A.B. Zamolodchikov, Nucl. Phys. B 241 (1984) 333.
\bibitem{DG} G. Delfino and P. Grinza, Nucl. Phys. B 682 (2004) 521.
\bibitem{NBRS} B. Nienhuis, A.N. Berker, E.K. Riedel and M. Schick, Phys. Rev. Lett. 43 (1979) 737.
\bibitem{LW} E.H. Lieb and F.Y. Wu, in Phase Transitions and Critical Phenomena, edited by C. Domb and M.S. Green (Academic, New York, 1972), Vol. 1, p. 331.
\bibitem{BH} J.K. Burton, Jr. and C.L. Henley, J. Phys. A 30 (1997) 8385.
\bibitem{AF1} G. Delfino, J. Phys. A 34 (2001) L311. 
\bibitem{AF2} G. Delfino, in Statistical Field Theories, NATO Science Series II, vol.~73, p.~3, Kluwer Academic Publishers, 2002. 
\bibitem{CJS} J. Cardy, J.L. Jacobsen and A.D. Sokal, J. Stat. Phys. 105 (2001) 25.
\bibitem{OO} H. Otsuka and Y. Okabe, Phys. Rev. Lett. 93 (2004) 120601.
\bibitem{FZ_flows} V.A. Fateev and Al.B. Zamolodchikov, Phys. Lett. B 271 (1991) 91.
\bibitem{ZF_paraf} A.B. Zamolodchikov and V.A. Fateev, Sov. Phys. JETP 62 (1985) 215.
\bibitem{Saleur} H. Saleur, Nucl. Phys. B360 (1991) 219.
\bibitem{JS} J.L. Jacobsen and H. Saleur, Nucl. Phys. B743 (2006) 207.
\bibitem{Ikhlef} Y. Ikhlef, Mod. Phys. Lett. B 25 (2011) 291.
\bibitem{Huang} Y. Huang, K. Chen, Y. Deng, J.L. Jacobsen, R. Koteck\'y, J. Salas, A.D. Sokal and J.M. Swart, Phys. Rev. E 87 (2013) 012136.
\bibitem{DHJSS} Y. Deng, Y. Huang, J.L. Jacobsen, J. Salas, and A.D. Sokal, Phys. Rev. Lett. 107 (2011) 150601.



\end{thebibliography}
\end{document}